\documentstyle[12pt,epsfig]{article}
\begin{document}
\noindent
\begin{center}
{\Large {\bf Probing Scalar-Photon Coupling in the Early Universe: Implications for CMB Temperature and Anisotropies  }}\\
\vspace{2cm}
 ${\bf Yousef~Bisabr}$\footnote{e-mail:~y-bisabr@sru.ac.ir.}\\
\vspace{.5cm} {\small{Department of Physics, Shahid Rajaee Teacher
Training University,
Lavizan, Tehran 16788, Iran}}\\
\end{center}
\begin{abstract}
The Hubble tension, as a persistent discrepancy between early-time and late-time measurements of the Hubble constant, motivates explorations of new physics in the early 
Universe. In a recent early dark energy (EDE) model, we introduced a scalar 
field interacting with the radiation sector at early-time before recombination. We showed that 
such a scalar-photon coupling can lead to an 
accelerated expansion phase in which the energy density of scalar component dilutes faster than 
 radiation does, a crucial feature for a successful EDE model. In the present work, we extend 
our analysis to investigate how this scalar-photon coupling affects the CMB temperature-redshift law and CMB anisotropies. We demonstrate that the 
temperature-redshift law deviates from the standard relation $T(z)\propto (1+z)$  due to the 
scalar-photon coupling. This deviation is controlled by a model parameter $\epsilon$, which quantifies the rate of energy transfer between the scalar field and radiation. We also argue that a positive value of $\epsilon$ shifts the acoustic peaks to larger scales, which potentially alleviates the Hubble tension. 
These findings suggest that scalar-photon coupling is a testable mechanism for reconciling 
different cosmological observations.

\end{abstract}

~~~~~~~~~~~~~~~~~~~~~~~~~~~~~~~~~~~~~~~~~~~~~~~~~~~~~~~~~~~~~~~~~~~~~~~~~~~~~~~~~~~~~~~~~~~~~~~~~~~~~
\section{Introduction}
The CMB serves as a powerful probe of the early Universe, capturing thermal radiation from the recombination era at redshift 
$z\approx 1100$. In the standard $\Lambda$CDM model, the CMB temperature is given by the adiabatic relation $T(z)=T_0(1+z)$ with $T_0\approx 2.725~K$
 being the present-day value, reflects a Universe that evolves through radiation, matter, and dark energy domination \cite{peeb}. CMB anisotropies, encoded in the temperature power spectrum $C_l$, reveal primordial perturbations, acoustic oscillations in the photon-baryon fluid and the curvature of spacetime, whose characteristic peaks tightly constrain cosmological parameters \cite{hu}\cite{planck}. Any deviation from this standard model, either in the temperature-redshift relation or the anisotropies spectrum, signal new physics beyond $\Lambda$CDM which motivates  studies of alternative cosmological evolution.\\
 Scalar fields have long underpinned cosmological extensions that elucidate phenomena from inflation to dark energy. During the radiation-dominated era, they are typically subdominant and their energy density rapidly diluted by radiation-driven expansion. Scalar-radiation couplings affect this dilution, background evolution and perturbation dynamics in the early Universe. Initial applications of interactions emerge in inflationary reheating where scalar fields like the inflaton decayed into radiation to trigger the hot Big Bang era \cite{kof}. Subsequent researches explored interactions between scalar fields and photons or other Standard Model fields, often inspired by string theory or axion-like fields, impacting cosmological observables such as the CMB \cite{carroll}\cite{kap}. Recently, scalar-photon interactions have been invoked to resolve anomalies such as the Hubble tension, leveraging energy exchange between dark sectors and radiation \cite{val}\cite{sch}\cite{bis0}.\\
 In our previous work \cite{bis1}, we demonstrated that a scalar field endowed with a potential and exponentially coupled to radiation during the radiation-dominated era alters the expansion rate of the Universe. This scalar-photon coupling generates an effective fluid whose equation of state mimics a cosmological constant which offers a novel mechanism by which the background cosmology is altered. Building on this, the current study investigates the implications of such a scalar-photon coupling for two key CMB observables: the temperature-redshift relation and the anisotropies. In standard adiabatic expansion, a non-interacting radiation fluid yields $T(z)\propto (1+z)$.
 The introduction of scalar-photon coupling causes energy exchange with the possibility that $T(z)$
 differs from the usual expression. Similar deviations, studied in decaying dark energy or photon dissipation context \cite{lima2}\cite{oph} face tight constraints from CMB spectral distortions and high-redshift temperature observations \cite{fix}\cite{av}. In addition, the coupling influences the evolution of density perturbations, distorting the CMB anisotropy power spectrum $C_l$ by shifting acoustic oscillations and damping scales, effects similar to those in models with varying constants or non-standard recombination \cite{hart}\cite{gali}.\\
The motivation for this study is twofold. First, deviations from the CMB temperature-redshift law provides a new window onto evolution of early Universe, enhancing constraints from galaxy surveys and supernova observations. Second, shifts in CMB anisotropy spectra, such as positions or heights of the acoustic peaks, produce detectable imprints in high-precision data from Planck \cite{planck} and future experiments like the Simons Observatory \cite{ade}. Based on our earlier work on background evolution, we seek to quantify how scalar-photon coupling modifies these CMB features, delivering testable predictions for non-standard cosmological models.\\  
 This paper is organized as follows: In Section 2, we give an overview of the theoretical framework established in our previous work where we introduced a scalar-photon coupling and investigated its implications for the background evolution of the Universe. We show that the resulting effective fluid from this interaction behaves as a cosmological constant offering a potential resolution to the Hubble tension. In Section 3, we extend our analysis to derive the modified temperature-redshift relation for the CMB. We will show that the standard temperature-redshift relation is modified and the modification is controlled by the rate of energy exchange between the scalar field and photons. In Section 4, we discuss implications of the scalar-photon coupling for the CMB anisotropies. We also deal with how the interaction affects the power spectrum and acoustic features. Finally, in Section 5, we combine these results discussing their implications for cosmological observations and presenting our result in a brief overview.

~~~~~~~~~~~~~~~~~~~~~~~~~~~~~~~~~~~~~~~~~~~~~~~~~~~~~~~~~~~~~~~~~~~~~~~~~~~~~~~~~~~~~~~~~~~~~~~~~~~~~~~~~~~~~~~~~~~~~~~~~~~~~~~~
\section{Interacting EDE, revisited}
We start with the action functional:
\begin{equation}
S= \int d^{4}x \sqrt{-g} \{\frac{1}{2}R
-\frac{1}{2}g^{\mu\nu}\nabla_{\mu}\phi
\nabla_{\nu}\phi-V(\phi)+C(\phi)L_{m}\}
\label{a1}\end{equation}
where $R$ is the Ricci scalar and $g$ is the determinant of the metric $g_{\mu\nu}$.
The action shows that a minimally coupled scalar field $\phi$ with the potential $V(\phi)$ is coupled with the matter Lagrangian density $L_m$ through the coupling function $C(\phi)$. We would like to study the implications of this action in the early Universe, particularly during the radiation-dominated era before recombination. This epoch is characterized by the dominance of radiation over other components. Consequently, we take the matter Lagrangian $L_m$ to represent a radiation fluid with energy density $\rho_{\gamma}$ and pressure $p_{\gamma}$. We also adopt an exponential coupling function $C(\phi)=e^{-\sigma\phi}$ with $\sigma$ being a coupling parameter\footnote{In Einstein frame representation of BD theory, this parameter is related to the BD parameter \cite{bis0}.}.
This framework allows us to examine how the scalar field $\phi$ interacts with photons in the early Universe and to analyze energy exchange between $\phi$ and the radiation fluid.  For a spatially flat Friedmann-Robertson-Walker (FRW) metric, (\ref{a1}) gives the gravitational equations
\begin{equation}
3H^2= \rho_{eff}= e^{-\sigma\phi}\rho_{\gamma}+\rho_{\phi}
\label{a4}\end{equation}
\begin{equation}
2\dot{H}+3H^2= -p_{eff}=-e^{-\sigma\phi}p_{\gamma}-p_{\phi}
\label{a5}\end{equation}
where $\rho_{\phi}=\frac{1}{2}\dot{\phi}+V(\phi)$ and $p_{\phi}=\frac{1}{2}\dot{\phi}-V(\phi)$ with $\omega_{\phi}=\frac{\rho_{\phi}}{p_{\phi}}$.
The radiation sector is not conserved due to coupling with $\phi$ and the (non-)conservation equations are \footnote{We have chosen $L_m=p_m$ for the Lagrangian density.},
\begin{equation}
\dot{\rho}_{\phi}+3H(\omega_{\phi}+1)\rho_{\phi}=-\frac{1}{3}\sigma e^{-\sigma\phi}\dot{\phi}\rho_{\gamma}
\label{a2}\end{equation}
\begin{equation}
\dot{\rho}_{\gamma}+4H\rho_{\gamma}=Q \label{a3}\end{equation}
where $Q\equiv \frac{4}{3}\sigma \dot{\phi}\rho_{\gamma}$. The equation (\ref{a3}) is equivalent to
\begin{equation}
\frac{d\rho_{\gamma}}{\rho_{\gamma}}+4\frac{da}{a}=\frac{4}{3}\sigma d\phi
\label{aa6}\end{equation}
which has the solution
\begin{equation}
\rho_{\gamma}=\rho_{0\gamma}a^{-4} e^{\frac{4}{3}\sigma\phi}
\label{ba6}\end{equation}
with $\rho_{0\gamma}$ being an integration constant. This can be recasted as
\begin{equation}
\rho_{\gamma}=\rho_{0\gamma}a^{-4+\epsilon}
\label{a6}\end{equation} with
\begin{equation}
\epsilon\equiv \frac{4\sigma\phi}{3\ln a}\label{a16}\end{equation}
 The coupling between the scalar field and the radiation component is governed by the coupling parameter $\epsilon$� which� determines the strength and direction of energy transfer between the two sectors.  For $\epsilon>0$, energy flows from $\phi$ into the radiation field leading to the production of radiation. In this case, the energy density $\rho_{\gamma}$ decreases more slowly compared to the standard adiabatic evolution where $\rho_{\gamma}\propto a^{-4}$. Conversely, when $\epsilon<0$ energy is transferred from radiation into the scalar field resulting in the depletion of radiation. This process causes $\rho_{\gamma}$ to dilute faster than the conventional $a^{-4}$ scaling. From the equation (\ref{a16}), we observe that $\epsilon$ is generally a dynamical function of $\phi$ and the scale factor $a$. However, for simplicity and analytical tractability, we treat $\epsilon$ 
as a constant parameter. While this assumption ignores its potential time dependence, it remains valid in scenarios where $\epsilon$ is small ensuring minimal deviations from the standard evolution. Under the approximation of a constant $\epsilon$, (\ref{a16}) implies that the scalar field $\phi$  varies logarithmically with the scale factor 
\begin{equation}
\phi=\gamma\ln a
\label{a7}\end{equation}
where $\gamma\equiv \frac{3\epsilon}{4\sigma}$. This behavior follows directly from the scalar-photon coupling and leads to the relation $\dot{\phi}=\gamma H$ implying that the rate of change of $\phi$� is essentially proportional to the Hubble parameter.\\ By using equations (\ref{a4}) and (\ref{a5}), we have the equation of state parameter of the effective fluid defined as
\begin{equation}
\omega_{eff}\equiv \frac{p_{eff}}{\rho_{eff}}
=\frac{ e^{-\sigma\phi}p_{\gamma}+p_{\phi}}{e^{-\sigma\phi}\rho_{\gamma}+\rho_{\phi}}
\label{a11}\end{equation}
Combining the latter with (\ref{a4}) and (\ref{a7}) gives
\begin{equation}
\omega_{eff}=(1-\frac{1}{3}\gamma^2)\frac{\frac{1}{3}e^{-\sigma\phi}\rho_{\gamma}-V(\phi)}{e^{-\sigma\phi}\rho_{\gamma}+V(\phi)}
\label{a12a}\end{equation}
When the scale factor $a$ approaches zero in the early Universe, the scalar field $\phi$ goes to infinity as inferred from (\ref{a7}). This causes the term $e^{-\sigma\phi}\rho_{\gamma}$� in (\ref{a12a}) to vanish, leading the effective equation of state $\omega_{eff}$ to stabilize at a constant value: $\omega_{eff}\rightarrow\frac{1}{3}\gamma^2-1$. For $\gamma<1$, this parameter is confined to the range $-1<\omega_{eff}<0$. More specifically, when $\gamma<<1$ (or equivalently, $\epsilon <<\frac{4}{3}\sigma$), $\omega_{eff}$ approaches $-1$ indicating that the effective fluid mimics behavior of a cosmological constant. It should be emphasized that this result holds independently of the scalar field potential $V(\phi)$ or the equation of state parameter of the scalar field itself.\\
For the scalar field to serve as a viable EDE candidate, its energy density must redshift faster than radiation as the Universe expands. This ensures us that the contribution of $\phi$
becomes negligible prior to recombination. Such behavior preserves the standard evolution of structure formation and is consistent with observational constraints from the CMB and large-scale structure.
To demonstrate this behavior, we solve Equation (4) under the assumption that $\omega_{\phi}\equiv\alpha\approx~const$. This simplification enables us to express the scalar field energy density $\rho_{\phi}$ as a function of the scale factor. In this case, Equation (4) can be reformulated as
 \begin{equation}
\dot{r}+[(\epsilon-4)+3(\alpha+1)] Hr=-\frac{\epsilon}{4} H a^{-\frac{3\epsilon}{4}}
\label{a8}\end{equation}
where (\ref{a7}) is used and $r\equiv\frac{\rho_{\phi}}{\rho_{\gamma}}$  measures the fractional energy densities of $\phi$
and radiation. The equation (\ref{a8}) can be rewritten as
\begin{equation}
\frac{dr}{da}+[(\epsilon-4)+3(\alpha+1)]\frac{r}{a}=-\frac{\epsilon}{4} a^{-\frac{3\epsilon}{4}-1}
\label{a8-1}\end{equation}
where the solution is
\begin{equation}
r(a)=\frac{\epsilon}{4\lambda}a^{-\frac{3\epsilon}{4}}
+C a^{-\lambda-\frac{3\epsilon}{4}}
\label{a9}\end{equation}
with $\lambda\equiv(3\alpha-1)+\frac{\epsilon}{4}$ and $C$ being an integration constant. Imposing a boundary condition at some early scale factor $a_c$� (just after inflation ends) allows us to fix the constant $C$�. More precisely, we demand that as the scale factor approaches this critical value, the function $r$�  satisfies the condition  $r(r\rightarrow a_c)\rightarrow 1$. It corresponds to a cosmological scenario where the energy densities of the scalar field and radiation are equal at $a=a_c$ ensuring a balance between the two as contributors to the total energy density during this primordial stage of cosmic evolution. This leads to the expression 
\begin{equation}
r(a)=\frac{\epsilon}{4\lambda}a^{-\frac{3\epsilon}{4}}\{1+[\frac{4\lambda}{\epsilon}a^{\frac{3\epsilon}{4}}-(\frac{a}{a_c})^{\frac{3\epsilon}{4}}
](\frac{a}{a_c})^{-\lambda-{\frac{3\epsilon}{4}}}\}
\label{a10}\end{equation}
When the scale factor $a$ significantly exceeds the critical value $a_c$ (i.e., $a>>a_c$), the term containing the coefficient $(\frac{a}{a_c})^{-\lambda-\frac{3\epsilon}{4}}$ approaches zero for positive values of 
$\lambda$ and $\epsilon$. This leads (\ref{a10}) to a simplified form $r(a)\approx \frac{\epsilon}{4\lambda}a^{-\frac{3\epsilon}{4}}$.
When $\epsilon>0$ (corresponding to the case that energy moves from $\phi$ to radiation) the ratio $r(a)$ decreases as the Universe expands. Physically, this behavior implies a progressive transfer of energy from $\phi$ to the radiation component, diminishing the relative contribution of the scalar field energy density $\rho_{\phi}$ compared to that of radiation $\rho_{\gamma}$ over time. This behavior is essential in cosmological models where the energy of the scalar field must reduce more quickly than that of radiation. This ensures that radiation becomes dominant before recombination aligning with the usual processes of structure formation in the Universe.
~~~~~~~~~~~~~~~~~~~~~~~~~~~~~~~~~~~~~~~~~~~~~~~~~~~~~~~~~~~~~~~~~~~~~~~~~~~~~~~~~~~~~~~~~~~~~~~~~~~~~~~~~~~~~~~~
\section{Modifying CMB temperature-redshift relation}
In this section, we derive a generalized temperature-redshift relation for the CMB by analyzing the effects of scalar-photon coupling on the evolution of the radiation fluid. This coupling introduces a non-adiabatic energy exchange that modifies the standard thermal evolution of the CMB. Central to our analysis are the conservation equation for the radiation sector, the equation (\ref{a3}), and the blackbody radiations which govern the photon energy density $\rho_{\gamma}$
and number density $n_r$ dynamics. \\
We use the equation (\ref{a3}) and rewrite it here 
$$
\dot{\rho}_{\gamma}+4H\rho_{\gamma}=Q
$$
where $Q=\frac{4}{3}\sigma\dot{\phi}\rho_{\gamma}$ accounts for energy exchange between the scalar field $\phi$ and photons due to their coupling. Unlike the standard adiabatic case ($Q=0$), where the photon energy density scales as $\rho_{\gamma}\propto a^{-4}$, the non-zero $Q$ indicates that energy is either injected into or extracted from the radiation sector depending on the sign of $Q$. In these cases, the evolution of the radiation energy density modifies according to (\ref{a6}). Furthermore, given that CMB exhibits the characteristics of blackbody radiation we employ the relations $\rho_{\gamma}\propto T^4$  and $n_{\gamma}\propto T^3$. These relations are fundamental to blackbody radiation reflecting the thermodynamic properties of a photon gas in thermal equilibrium, where the energy per photon scales as $T$ and the number of photons per unit volume scales as $T^3$.\\
Using these blackbody relations, we can connect the energy conservation equation (\ref{a3}) to the photon number density evolution. The number density $n_{\gamma}$ then evolves according to
\begin{equation}
\dot{n}_r+3n_rH=\frac{3n_r}{4\rho_{\gamma}}Q
\label{a13}\end{equation}
which follows directly from the equation (\ref{a3}) by noting that $\rho_{\gamma}\propto n_{\gamma}T$ 
and differentiating with respect to time. Physically, the equation (\ref{a13}) indicates that the photon number is not conserved when $Q\neq 0$ as the scalar-photon coupling either creates or annihilates photons depending on the sign of $Q$. To derive the temperature evolution, we combine $\rho_{\gamma}\propto T^{4}$ with (\ref{a3}) which gives 
\begin{equation}
\frac{\dot{T}}{T}+H=\frac{Q}{4\rho_{\gamma}}
\label{a12}\end{equation}
This differential equation governs the rate of change of the CMB temperature $�T(z)$. Integrating gives
 \begin{equation}
 T(z)=T_0(1+z)^{1-\beta}
\label{a15}\end{equation}
with $\beta=\epsilon/4$. This equation demonstrates that the CMB temperature evolves differently from the standard adiabatic law $T(z)=T_0(1+z)$ due to the scalar-photon coupling. For $\beta>0$, corresponding to energy transfer from the scalar field $\phi$ to radiation ($\epsilon>0$), the temperature redshifts more slowly, due to photon creation opposing the standard redshift dilution. 
This slower redshift corresponds to a higher photon number density than expected, as seen in the equation (\ref{a13}), and could manifest as subtle changes in CMB observables such as the thermal Sunyaev-Zeldovich effect or high-redshift temperature measurements.
Conversely, $\beta<0$ (energy injection into $\phi$) increases the temperature decline due to photon annihilation. This case implies a reduced photon number density potentially affecting the CMB power spectrum or spectral distortions. This asymmetry, driven by the interaction term $Q$ affects the thermal history and potentially the CMB anisotropy spectrum. \\
The functional form of $T(z)$ given by (\ref{a15}) parallels with some modifications reported in the literature. For instance, Lima et al., derived  similar relation from vacuum or gravity-mediated decaying photons with $\beta$ acting as a photon creation parameter. They suggested $\beta\approx 0-0.20$ based on QSO absorption data though unconstrained statistically \cite{lima2}\cite{lima1}.  Avgoustidis et al., have considered  scalar-electromagnetic interaction (e.g., $F^{\mu\nu}F_{\mu\nu}e^{-\alpha\phi}$) in models with varying fundamental constants and dynamical dark
energy which yields the same $T(z)$ profile \cite{avg}. Jetzer et al., explored decaying dark energy models and found similar modified $T(z)$ relation. They used Sunyaev-Zeldovich and quasar absorption-line data yielding $\beta=0.027^{+0.04}_{-0.027}(68\% C.L.)$ \cite{jet}. Luzzi et al, analyze multi-frequency Sunyaev-Zeldovich data from $13$ clusters using intensity ratios and direct intensity methods to test $T(z)\propto (1+z)^{1-\alpha}$ and found $\alpha=0.024^{+0.068}_{-0.024}~(68\%C.L.)$ \cite{luzzi}. Fixsen et al. \cite{fix}, used COBE/FIRAS data to measure CMB spectral distortions. They constrained deviations from a perfect blackbody spectrum implying $|\beta|<0.01$ to avoid detectable distortions.

~~~~~~~~~~~~~~~~~~~~~~~~~~~~~~~~~~~~~~~~~~~~~~~~~~~~~~~~~~~~~~~~~~~~~~~~~~~~~~~~~~~~~~~~~~~~~~~~~~~~~~~~~~~~~~~~~~~~~~~~~~~~~~~~~~~~~~~~~
\section{Impacts of scalar-photon coupling on CMB anisotropies}
Anisotropies of the CMB offer a snapshot of the behavior of the early Universe driven by acoustic oscillations during the radiation-dominated era before recombination. In this section, we present a brief qualitative study of effects of the scalar-photon coupling, described by the action (\ref{a1}), on the CMB anisotropies. In this context, the effective fluid consists of photons and $\phi$ which their coupling alters the photon energy density evolution from $\rho_{\gamma}\propto a^{-4}$ to $\rho_{\gamma}\propto a^{-4+\epsilon}$ where $\epsilon$ measures energy transfer between the two components. Comparing the uncoupled ($\epsilon=0$) and the coupled ($\epsilon\neq 0$) cases, we explore how this interaction affects the speed of sound, sound horizon and ultimately CMB power spectrum and thus gain insight into fixing the Hubble tension.\\
In the uncoupled case ($\epsilon=0$), the early Universe follows the usual patterns in $\Lambda$CDM. The radiation energy density decreases according to the standard rule $\rho_{\gamma}\propto a^{-4}$. Since baryons and $\phi$ remain subdominant before recombination and radiation essentially dominates the effective fluid, the sound speed is given by
$c_s\approx\frac{1}{\sqrt{3}}$ which is typical for something behaving like a relativistic gas. This sound speed characterizes the sound horizon, namely the distance that acoustic waves can travel by the time of recombination, which sets the characteristic scale of the CMB acoustic peaks. This distance is essential because it determines the scale of the acoustic peaks which we observe in the CMB. The power spectrum exhibits a series of peaks at multipoles $l$, which are related to the ratio between the angular diameter distance and the sound horizon. The peak amplitudes tell us about the strength of gravitational driving and the resulting acoustic oscillations while a damping tail appears at higher multipoles due to photon diffusion. This framework establishes the baseline anisotropy pattern commonly observed in the standard CMB analyses such as those conducted by Planck \cite{planck}.\\
In the coupled case ($\epsilon\neq 0$), on the other hand, the scalar-photon interaction leads to significant changes in dynamics of the effective fluid. The radiation energy density  evolves according to $\rho_{\gamma}\propto a^{-4+\epsilon}$ and the overall energy density is given by $\rho_{eff}\propto a^{-4+\epsilon/4}$. The latter relation indicates how both the photons and the scalar field contribute to the total energy. Since baryons are subdominant, the dynamics really depends on how photons and $\phi$ interact. When $\epsilon>0$, the slower dilution compared to $a^{-4}$ effectively increases radiation energy density at early times. This raises the sound speed above the value $\frac{1}{\sqrt{3}}$ by amplifying the pressure response to density perturbations. Conversely, $\epsilon<0$ means faster dilution implying a lower energy density and decreasing the sound speed. These shifts in sound speed affect the sound horizon: a higher $c_s$ ($\epsilon>0$) allows sound waves to travel farther by recombination, while a lower $c_s$ ($\epsilon<0$) limits their range altering the scale of acoustic oscillations imprinted on the CMB.\\
Comparing the two cases $\epsilon=0$ and $\epsilon\neq 0$, reveals distinct effects on CMB anisotropies: 
In the uncoupled case, the sound speed and the sound horizon produce a fixed set of acoustic peaks with multipoles determined by a standard sound horizon while the damping tail is shaped by diffusion in a radiation-dominated fluid. In the coupled case and $\epsilon>0$, the higher sound speed enlarges the sound horizon which shifts the acoustic peaks to lower $l$ values corresponding to larger angular scales. This shift could suggest a lower Hubble constant inferred from the CMB potentially aligning early Universe cosmology with late-time measurements and easing the Hubble tension. The stiffer fluid leads to raising peak amplitudes by enhancing oscillation strength while the damping tail becomes less pronounced as a higher sound speed reduces the impact of diffusion. 
For $\epsilon<0$, the lowered sound speed reduces the sound horizon driving peaks to higher $l$ values (smaller angular scales) which might lower peak amplitudes due to weaker oscillations and steepen the damping tail as diffusion effects intensify. \\
Thus the scalar-photon coupling introduces a tunable parameter $\epsilon$ which adjusts the CMB anisotropies in ways that go beyond the standard $\epsilon=0$ scenario. The case $\epsilon>0$ supports an early acceleration phase which aligns with our previous findings \cite{bis1}. This may potentially help to address the Hubble tension by shifting the peak positions. On the other hand, the case $\epsilon<0$ could make things more difficult and $\epsilon$ should be kept well-constrained. The impact on sound speed and sound horizon, influenced by the scalar-photon effective fluid, emphasizes how this coupling can leave distinct marks on the CMB. These unique signatures can be tested against precise data from Planck or future missions like the Simons Observatory, adding to our analysis of the temperature-redshift relationship.

~~~~~~~~~~~~~~~~~~~~~~~~~~~~~~~~~~~~~~~~~~~~~~~~~~~~~~~~~~~~~~~~~~~~~~~~~~~~~~~~~~~~~~~~~~~~~~~~~~~~~~~~~~~~~~~~~~~~~~~~~~~~~~~~~~~~~~
\section{Conclusions}
In this work, we have investigated consequences of interaction between a scalar field $\phi$ and radiation in the radiation-dominated era. We study the impacts of this interaction on the CMB temperature-redshift relation and anisotropies. We have considered an exponential coupling function and demonstrated that the energy exchange between these components can significantly alter the thermal evolution of CMB and the dynamics of acoustic oscillations. Our key findings can be summarized as follows:\\
\textbf{Modified Temperature-Redshift Relation: }The scalar-photon coupling leads to a deviation from the standard relation $T(z)\propto (1+z)$ to $T(z)\propto (1+z)^{1-\frac{\epsilon}{4}}$ as a consequence of energy transfer between the scalar field $\phi$ and radiation. For $\epsilon>0$, energy is moving from $\phi$ to radiation leading to a slower temperature drop. On the other hand, $\epsilon<0$  speeds up decreasing of the CMB temperature. This non-adiabatic behavior provides a distinct signature that could be probed by high-precision CMB observations.\\
\textbf{Impact on CMB Anisotropies:} The scalar-photon coupling affects the sound speed and sound horizon which in turn changes the positions and heights of the acoustic peaks in the CMB power spectrum. When $\epsilon>0$, the sound horizon increases shifting the peaks to larger angular scales. This could help to alleviate the Hubble tension by aligning early- and late-time measurements of the Hubble constant. On the other hand, $\epsilon<0$ reduces the sound horizon shifting the peaks to smaller scales. These variations show how scalar-photon interactions can leave measurable imprints on the CMB anisotropy spectrum.\\
\textbf{Early Dark Energy (EDE) Behavior:} The scalar-photon interaction makes an effective fluid which behaves like a fluid with an equation of state close to $-1$.
This period of early accelerated expansion is transient, as $\rho_{\phi}$ dilutes faster than $\rho_{\gamma}$ , ensuring compatibility with standard recombination and structure formation.\\
Our findings indicate that scalar-photon coupling could be a viable mechanism to address some puzzling cosmological issues such as the Hubble tension while providing some testable predictions for upcoming CMB experiments including the Simons Observatory. The modified temperature-redshift relationship and the changes in the CMB power spectrum provide unique observational signatures that can put constraints on the coupling parameter $\epsilon$. Further studies could also explore how this coupling parameter affects other cosmological observables, such as large-scale structures giving us a deeper insight into the physics of the early Universe beyond the ΛCDM paradigm.

~~~~~~~~~~~~~~~~~~~~~~~~~~~~~~~~~~~~~~~~~~~~~~~~~~~~~~~~~~~~~~~~~~~~~~~~~~~~~~~~~~~~~~~~~~~~~~~~~~~~~~~~~~~~~~~~

~~~~~~~~~~~~~~~~~~~~~~~~~~~~~~~~~~~~~~~~~~~~~~~~~~~~~~~~~~~~~~~~~~~~~~~~~~~~~~~~~~~~~~~~~~~~~~~~~~~~~~~~~~~~~~~~~~~

\end{document}